\documentclass[conference,letterpaper]{IEEEtran}

\usepackage[utf8]{inputenc} 
\usepackage[T1]{fontenc}
\usepackage{url}
\usepackage{ifthen}
\usepackage{cite}

\usepackage{graphicx}
\usepackage{hyperref}
\usepackage{amsmath}
\usepackage{amssymb}
\usepackage{bm}
\usepackage{bbm}
\usepackage[ruled,vlined]{algorithm2e}
\usepackage{array}
\usepackage{booktabs}
\usepackage{multirow}
\usepackage{makecell}
\usepackage{mathtools}
\usepackage{amsthm}
\usepackage{float}

\begin{document}
\title{Causal (Progressive) Encoding over Binary Symmetric Channels with Noiseless Feedback}


\author{\IEEEauthorblockN{Amaael Antonini, Rita Gimelshein and Richard~D.~Wesel}
\IEEEauthorblockA{
Department of Electrical and Computer Engineering\\
University of California, Los Angeles, Los Angeles, CA 90095, USA\\
Email: amaael@g.ucla.edu, rgimelshein@g.ucla.edu, wesel@ucla.edu} 
}


\maketitle

\begin{abstract}
Traditional channel coding with feedback constructs and transmits a codeword only after all message bits are available at the transmitter. This paper joins Guo \& Kostina and  Lalitha et al. in developing approaches for causal (or progressive) encoding, where the transmitter may begin transmitting codeword symbols as soon as the first message bit arrives.  Building on the work of Horstein, Shayevitz and Feder, and Naghshvar \emph{et al.}, this paper extends our previous computationally efficient systematic algorithm for traditional posterior matching to produce a four-phase encoder that progressively encodes using only the message bits causally available.
Systematic codes work well with posterior matching on a channel with feedback, and they provide an immediate benefit when causal encoding is employed instead of traditional encoding.  Our algorithm captures additional gains in the interesting region where the transmission rate $\mu$ is higher than the rate $\lambda$ at which message bits become available. In this region, transmission of additional symbols beyond systematic bits, before a traditional encoder would have begun transmission, further improves performance.

\end{abstract}

{\let\thefootnote\relax\footnote{{ 
This research is supported by National Science Foundation (NSF) grant CCF-1955660. Any opinions, findings, and conclusions or recommendations expressed in this material are those of the author(s) and do not necessarily reflect views of the NSF.}}}

\section{Introduction}
\label{sec: problem setup}


Consider a communication system where information needs to be  communicated promptly and reliably (with a message error rate no more than $\epsilon$) from the source where it is generated across a binary symmetric channel (BSC) with feedback to a destination where it will be used, for example, in a control application like stabilizing a plant \cite{Sukhavasi}. 

Shannon \cite{Shannon1956} showed that feedback cannot improve the capacity of discrete memoryless channels, including the BSC. However, Burnashev \cite{Burnashev1976} showed that feedback can help increase the exponent with which the frame error rate (FER) decreased with blocklength, when using variable length coding scheme. Horstein developed a single phase transmission scheme that attains the capacity of the BSC using full feedback.  Other non-causal algorithms have shown to attain Burnashev's optimal error exponent, which include \cite{Schalkwijk1971,Schalkwijk1973,Tchamkerten2002,Tchamkerten2006,Naghshvar2012, Naghshvar2015}. Most recently, Lalitha \emph{et al.}\cite{lalitha2020causal} proposed a causal encoding version of Horstein's scheme over the BSC in the context of stabilizing a plant. 


This paper presents a practical causal encoder that extends the systematic approach of \cite{9174232} and studies its performance in the classical communication framework of communicating a $K$-bit message with a small, specified message error probability at the earliest possible decoding time. The \mbox{$K$-bit} arrive at the transmitter at a rate $\lambda$ [bits/second]. The channel allows transmission at a rate of $\mu$ [bits/second] over BSC($p$) with full feedback, where $p$ is the channel's crossover probability. Let the number of transmitted symbols required to decode the message be $\tau$, and the decoding time be $T_d$, where $T_d = \frac{\tau}{\mu}$ [seconds] for a causal system. Our main goal is to minimize $E[T_d]$. A secondary goal is to maximize the expected transmission rate $E[r]=E[\frac{K}{\tau}]$.

The primary contribution of this paper is an efficient encoding algorithm that considers the source data causally and achieves the rate performance of traditional posterior matching (PM) algorithms. This is made possible by a method to synthesize a larger-blocklength PM system from an existing shorter-blocklength PM system and new message information.  The new method maintains and sometimes reduces the computational complexity of the original PM system. 

Building on \cite{9174232}, 
a combined tree-list structure collects groups of messages with possibly different posterior probabilities while still preserving all the information about each item. Such trees have been used in coding before, e.g. by Schulman \cite{Schulman}, but we propose an approach to combine partially decoded systems into a single system. For this we use nodes that are also lists of cases from a smaller sub system. We also evaluate the regions where our algorithm would provide an advantage over other, non-causal encoding schemes. Additionally we show the properties of an optimal encoding rule and analyze the performance of the algorithm in terms of its ability to stay "close" to optimal encoding rule.

The rest of this paper is organized as follows: Sec. 
\ref{sec: notation} introduces the channel model.  Sec. \ref{sec:regions} considers systematic encoding and organizes the causal encoding problem according to regions of $\gamma=\lambda/\mu$. Sec. \ref{sec:SPM} reviews the systematic posterior matching (SPM) algorithm of \cite{9174232}.  Sec.  \ref{sec: causal algorithm} introduces our main algorithm, a causal encoding scheme capable of achieving the high performance of well known PM schemes like Burnashev \cite{Burnashev1975}, Nagshavar \cite{Naghshvar2015}, Horstein \cite{Horstein1963} and \cite{9174232}. Sec. \ref{sec:benefit} shows that equiprobable signaling  maximizes the expected increase in the posterior of the true message. Sec. \ref{sec:  simulation} shows simulation results of different scenarios using the causal encoding algorithm. Sec. \ref{sec: conclusion} provides concluding remarks.

\section{Channel Model}
\label{sec: notation}
\begin{figure}[t]
\centering
\includegraphics[width=0.45\textwidth]{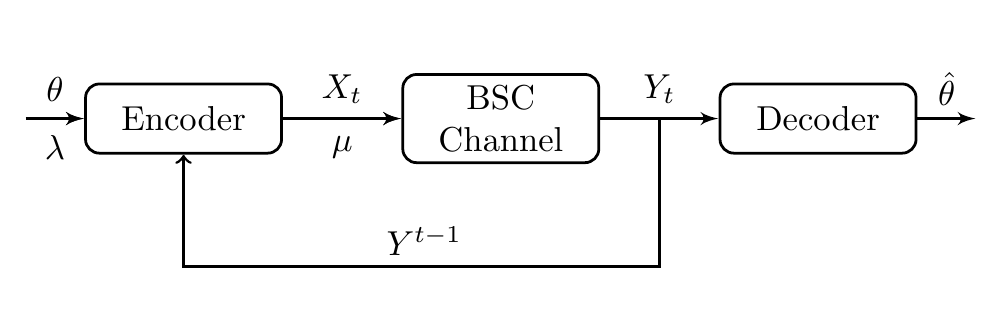}
\caption{System diagram of a DMC with full, noiseless feedback.}
\label{fig: system model}
\end{figure}

Figure \ref{fig: system model} shows a discrete memoryless channel (DMC) with full feedback, which has capacity C. For this paper, we restrict attention to the BSC. Let $t$ be the discrete time index $t = 1,2,\dots$. The $K$-bit true message is $\theta$, sampled from $U[\Omega] = U\{0,1\}^K$, and $\hat{\theta}$ denotes the estimate of $\theta$, with bits $\hat{b}^\theta_1, \dots, \hat{b}^\theta_{K}$. $X_t$ and $Y_t$ are variables representing the transmitted and received symbols, where $x_t$, $x_t$ denote a realization of these symbols.  The sequences $\{x_1, x_2,\dots, x_t\}$, $\{y_1, y_2, \dots,y^t\}$ are represented by $x^t$, $y^t$.
We seek a system that achieves a frame error rate (FER) $P(\hat{\theta} \neq \theta)$ that is less than $\epsilon$. Message bits become available at the transmitter at a rate of $\lambda$ [bits/second]. The link is able to transmit $\mu$ [bits/second] and their ratio is denoted by $\gamma = \frac{\lambda}{\mu}$.

\section{Systematic Encoding and Regions of $\lambda/\mu$}
\label{sec:regions}
In a traditional system, the transmitter waits until the entire $K$-bit message has been received before beginning to transmit. Such a system requires an expected transmission time no lower than $\frac{K}{\lambda} + \frac{K}{C\mu}$ seconds. Causal encoding permits transmission of $\frac{K}{\lambda}\mu$ symbols during the time the traditional system is waiting for the $K$ message bits to become available.  

A systematic PM algorithm, such as the one provided by \cite{9174232}, leads to a straightforward causal encoder, which we refer to as a systematically causal encoder (SCE),  where the $K$ systematic bits are transmitted as they become available and non-systematic bits are only transmitted after the entire message has been received. The expected decoding time for an SCE is lower bounded by assuming the transmission rate is $C$ as follows:
\begin{equation}
    E[T_d] \ge \max \left \{ \frac{K}{\lambda}, \frac{K}{\mu} \right \} + \frac{1}{\mu}\left (\frac{K}{C} - K \right )  
    \label{Old_bound}
\end{equation}

Using the systematic encoder of \cite{9174232}, the rate approaches the BSC capacity rapidly as $K$ becomes large, so the SCE bound of \eqref{Old_bound} is interesting for large $K$.  The bound separates the decoding time into two terms: the time until all $K$ systematic bits have been transmitted (left term) and the time required to transmit the subsequent non-systematic bits (right term).  The left term is characterized by two regions: one region where $\gamma=\lambda/\mu \ge 1 $ and another region in which  $\gamma < 1$.  

For the region where  $\gamma \ge 1$, using the algorithm proposed in \cite{9174232} would lead a decoding time $E[T_d]$ such that $E[\tau] \rightarrow \frac{K}{C}$ as $K$ becomes large.  Thus, for $\gamma \ge 1$, a "new" encoder is not needed for effective causal encoding using PM, since the encoder of \cite{9174232} may be applied directly.  However, when $\gamma < 1$,  $\frac{K}{\gamma} > K$ symbols (bits), may be transmitted by the time the full message becomes available, allowing for $K\left( \frac{1-\gamma}{\gamma} \right)$ non-systematic bits in addition to the $K$ systematic bits to be transmitted. This paper explores the use of causal (progressive) encoding to use all (or most) of the $\frac{K}{\gamma}$ available symbols to  reduce the expected decoding time $E[T_d]$. 


We note that the problem also simplifies as $\gamma \rightarrow 0$. For all but the last bit, one could simply repeat the bit and achieve the desired reliability. Specifically, for some $\gamma_1 << 1$, repetition of the systematic bits guarantees
\begin{equation}
    P(\hat{b}^\theta_i \neq b^\theta_i) \le 1- (1-\epsilon)^{1/K}
    ,\quad, i = 1,\dots, K-1\\
    \label{Low_BER}
\end{equation}
so that after all of the message bits have been received, all that remains is to repeat the final bit until it too has achieved 
\begin{equation}
    P(\hat{b}^\theta_K \neq b^\theta_K) \le 1- (1-\epsilon)^{1/K}
    \label{last_bit} \, .
\end{equation}
The region where another approach is necessary is then when $\gamma_1 < \gamma < 1$, and that is the focus of our paper.

\section{Systematic Posterior Matching}
\label{sec:SPM}

Algorithm \ref{alg: grouping}, shown below, is the systematic posterior matching (SPM) algorithm of \cite{9174232} and a fundamental building block of the proposed causal encoding algorithm.

\begin{algorithm}
\SetAlgoLined
\KwResult{index $\hat\theta$ s.t. $\rho_{\hat\theta}(\tau) \ge 1-\epsilon$}
\For{$t = 1, \dots, K$}{
    channel input: $x_t = b^\theta_t$, output: $y_t$\;
}
build sorted list of groups $\{\mathcal{G}_0,\dots,\mathcal{G}_K\}$ \; 
$\quad \mid \mathcal{G}_i\mid = \binom{K}{i}, \quad j \in \mathcal{G}_i \rightarrow \rho_j(K) = q^{K-i}p^i$ \;
 \While{$\underset{i}{\max}$ $\rho_i(t) < 1-\epsilon$}{
  build $S_0, S_1$ s.t. $S_0 \cup S_1 = \cup_i \mathcal{G}_i$, $S_0 \cap S_1 = \emptyset$, $|P(S_0) - P(S_1)|$ 'small'\;
  \textbf{if} needed: split $\mathcal{G}_i$ into  $\mathcal{G}^1_i, \quad \mathcal{G}^2_i$ \;
  
  \textbf{if }$\theta \in S_0$ \textbf{then}  $x_{t+1} = 0$, \textbf{else }$x_{t+1} = 1$ \textbf{end}\;

  channel input: $x_{t+1}$, output: $y_{t+1}$\;
  $j \in \mathcal{G}_i \in S_{y_{t+1}}\rightarrow \rho_j(t+1) = \frac{\rho_j(t)q}{P(S_{y_{t+1}})(q-p) + p}$\;
  $j \in \mathcal{G}_i \notin S_{y_{t+1}} \rightarrow \rho_j(t+1) = \frac{\rho_j(t)p}{P(S_{y_{t+1}})(p-q) + q}$ \;
  set $t = t+1$\;
  merge $S_0$ and $S_1$ into $\{\mathcal{G}_0,\dots\}$
 }
 \caption{Algorithm from \cite{9174232}}
 \label{alg: grouping}
\end{algorithm}

SPM begins by transmitting the $K$ message bits systematically.  Then, the possible received messages are sorted at the receiver into $K+1$ initial groups, where each message in a group shares the same Hamming distance from the received word and hence the same posterior.

PM updates are performed by constructing sets $S_0$ and $S_1$ on the groups so that $|P(S_0) - P(S_1)|$ is small, splitting groups when necessary.  As described  further in \cite{9174232}, groups with low posteriors can even be merged to provide further complexity reduction.

 \section{The Sub-Block-Combining SPM Algorithm}
 \label{sec: causal algorithm}
The sub-block-combining SPM (SBC-SPM) algorithm allows transmission of non-systematic bits before the full message has been received by performing SPM on sub-blocks. For each sub-block, Algorithm \ref{alg: grouping} places each message segment of size $L_i$ in a list of groups along with its associated posterior probability  $P(j = B^\theta_i)$.

\begin{figure}[t]
\centering
\includegraphics[width=0.45\textwidth]{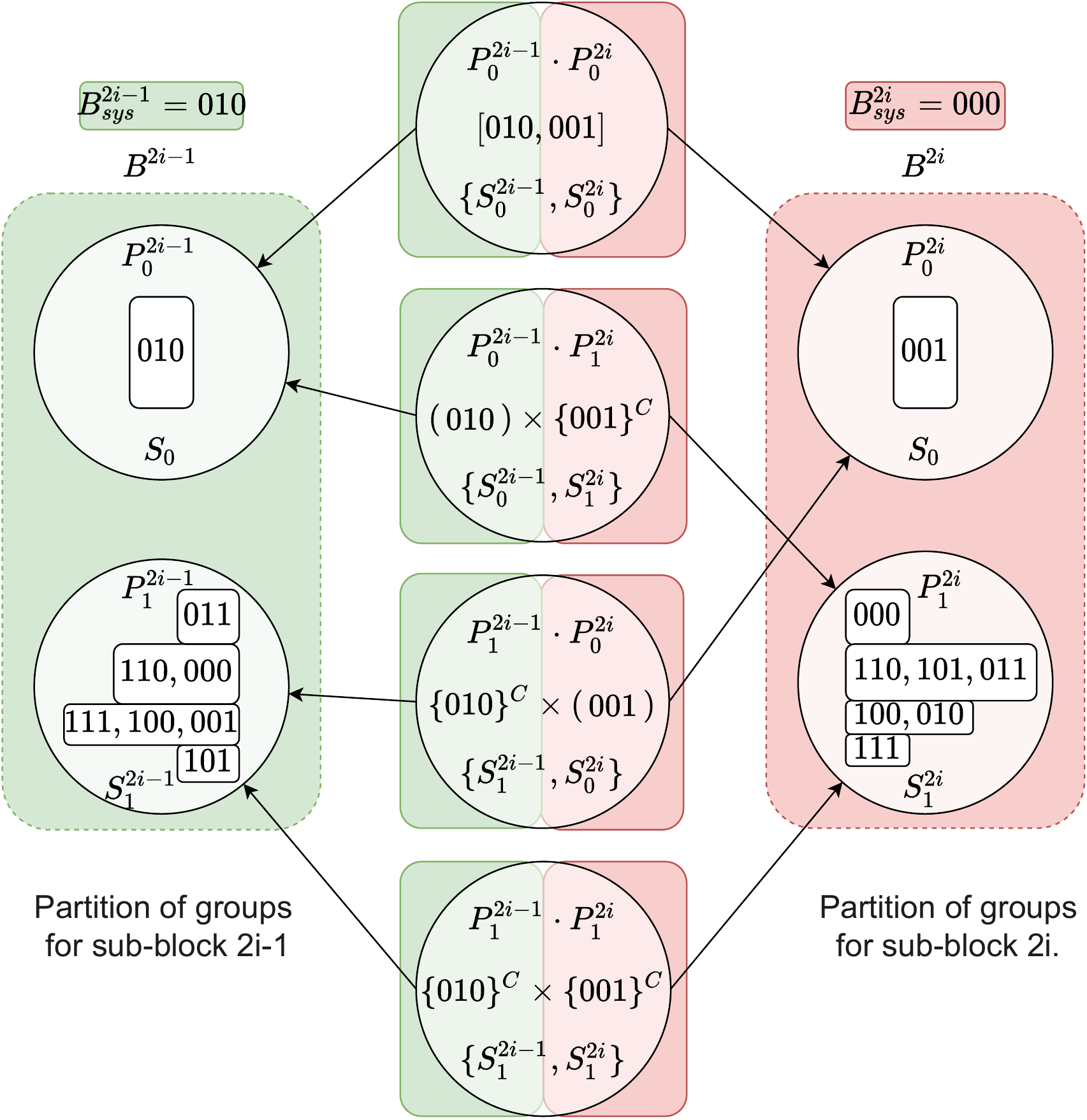}
\caption{Construction of tree nodes, in the middle, from 2 sub-blocks 2i-1, 2i, in left and right. The sub-blocks are at the leaf level and the new nodes are one level above. In the example, the sub-blocks are 3 bits each, and sample systematic bits are shown, as well as a possible system progress at the time of tree construction. It can be verified that the 4 new nodes encompass all possible combinations of the two sub-blocks into a new 6-bit block.}

\label{fig: size 4 tree}
\end{figure}
 \subsection{Initialization and Separate SPM Encoding of Sub-blocks}

SBC-SPM segments the $K$-bit message into $N$ sub-blocks $\{B_1,\dots,B_N\}$ of lengths $L_i,\dots,L_N$, where N is a power of 2. When bits of sub-block $B_i$ are available at the transmitter, SPM uses Algorithm \ref{alg: grouping} to produce symbols that encode $B_i$.  

\begin{figure}[t]
\centering
\includegraphics[width=0.45\textwidth]{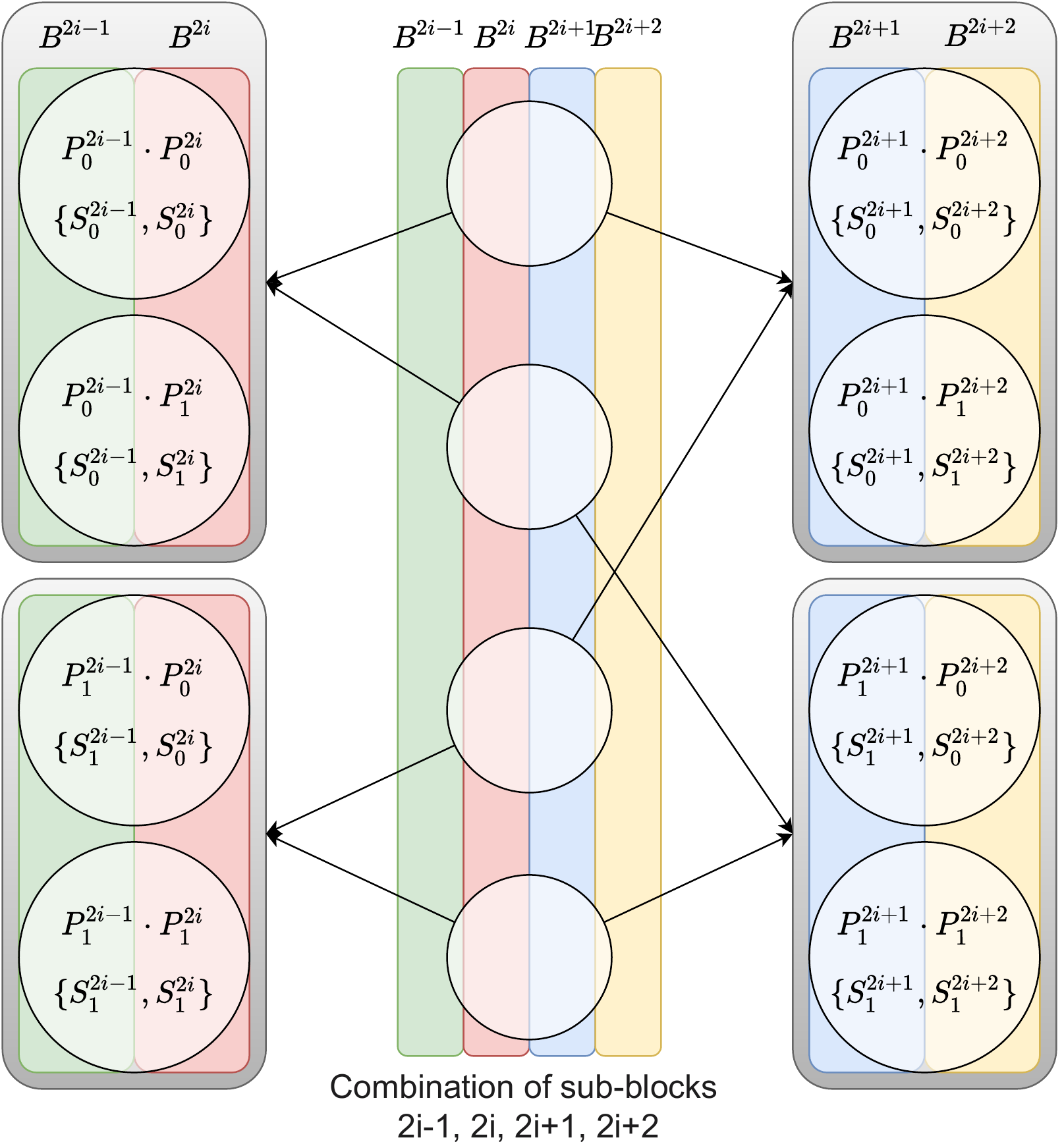}
\caption{Construction of higher level nodes, in the middle, from lower level nodes that are not leaves. The example shows the new nodes 2 levels above the leaves, and collect all possible combinations of the 4 sub-blocks 2i-1, 2i, 2i+1, 2i+2. }
\label{fig: nodes from nodes}
\end{figure}
\subsection{Combining the Separate Sub-Blocks}

Once all sub-blocks have been initially transmitted using Algorithm \ref{alg: grouping}, the sub-blocks $B_1,\dots,B_N$ are combined into a single block that contains the entire message.  The combination process places each $K$-bit message in a binary tree structure where each final leaf is a group of segments from a single sub-block.  The overall structure has a linked list of four nodes at the top level.  Each of these four nodes contains a subset of possible $K$-bit messages stored in a binary tree.  These four subsets are mutually exclusive and collectively exhaustive.  

Each node in one of the four binary trees has two branches that point respectively to nodes below that represent the first half and the second half of the sequences represented by the node.  To reconstruct a message with eight sub-blocks the tree is traversed down three levels identifying eight paths, one for each sub-block. For each of these paths, there are binary decisions at each node to selecting which of two subsets contains the desired sub-string of the overall message.

The trees are built from the bottom up, starting with $2^{N-1}$ trees, each representing all possible combinations of a pair of sub-blocks.  Fig. \ref{fig: size 4 tree} shows a tree with four level-1 nodes containing all combinations of sub-blocks  $B_{2i -1}$ and $ B_{2i}$.  For this example, sub-blocks contain three-bit sequences. After Algorithm \ref{alg: grouping}, each three-bit sequence is either in $S_0$ or $S_1$ for that sub-block. To combine the two sub-blocks each six bit message is assigned to one of four level-1 nodes depending on whether each three-bit string resides in  $S_0$ or $S_1$.

Fig. \ref{fig: nodes from nodes} shows how the four level-1 nodes created from two sub-blocks in the example of Fig. \ref{fig: size 4 tree}  are combined with four level-1 nodes from combining two other sub-blocks to produce four level-2 nodes. Each group of four level-1 nodes is partitioned into two sets of level-1 nodes.  Then the four level-two nodes are created corresponding to the four possible combinations of these sets of pairs level-1 nodes.  Each node has two branches, one pointing to a pair of level-1 nodes for two sub-blocks from Fig. \ref{fig: size 4 tree}  and the other pointing to a pair of level-1 nodes for the other two sub-blocks.

This process continues until all sub-blocks have been merged into a single block.  Four nodes are created at each level, and the combination process concludes with four nodes, each sitting at the top of a binary tree.  This process computes the correct posterior for each $K$-bit. The tree completed structure is used to perform SPM according to Algorithm  \ref{alg: grouping} on the overall message until its posterior exceeds $1-\epsilon$.


\subsection{Complexity}
\label{sec:complexity}
The computational complexity of this algorithm depends on the number of transmissions needed, which is linear in $K$, and the number of operations per transmission. \mbox{SBC-}SPM as well as SPM, partition, update and merge their lists of trees and groups respectively. These require visiting each item at most once.   We have shown in \cite{9174232} that the list size in SPM grows linearly in $K$.
The SBC-SPM algorithm compactly collects in four trees all the items in the product set $B_1 \times B_2 \times \cdots \times B_N$, each a group of lexicographically consecutive messages with shared posterior probabilities.  The list grows in the same manner as in SPM. However, multiple split operations might be required to obtain a desired partition of the list.  The number of splits depends on tree depth and branch state, and is upper bounded by $\log_2(N)$. The list in SBC-SPM is then of order $O(K \log_2(N))$. Two natural choices for $N$ are constant and linear in $K$. The total number of split operations is of order  $O(K \log^2_2(N))$, and the overall computational complexity is of order $O(K^2 \log_2(N))$. If $N$ is linear in $K$, the complexity becomes $O(K^2 \log_2(K))$.
Finally, this algorithm exhibits the same properties that allowed the SPM algorithm in \cite{9174232} truncate the list it operates on and obtain a much  lower computational complexity.



\subsection{Benefit of Combining the Sub-Blocks}
\label{sec:benefit}
Combining  is important for efficiency because PM encoding of a single sub-block of the message becomes inefficient if it continues for too many symbols. To explain this, we first show that partitioning the overall message into approximately equally likely sets $S_0$ and $S_1$ maximizes efficiency.   

Let $\rho_\theta(t)$ be the posterior corresponding to the true message $\theta$. The goal of SPM encoding is to increase $\rho_\theta(\tau)$ to $1-\epsilon$ at the smallest decoding time $T_d$. To that end, at each time $t$, we seek to maximize the expected increase in $\rho_\theta(t)$: $E[\frac{\rho_\theta(t+1)}{\rho_\theta(t)}]$. 

To  maximize $E[\frac{\rho_\theta(t+1)}{\rho_\theta(t)}]$, SPM should group the messages so that the sets $S_0(t)$ and $S_1(t)$ are approximately equally likely.  The weight update $w_\theta(t)$  scales the posterior $\rho_\theta$ as follows: $\rho_\theta(t+1) = w_\theta(t) \rho_\theta(t)$.  The value of  $\rho_\theta(t)$ is fixed before each transmission, so maximizing $E[\frac{\rho_\theta(t+1)}{\rho_\theta(t)}]$ is equivalent to maximizing the expected update $E[w_\theta(t)]$.
\begin{figure}[t]
\centering
\includegraphics[width=0.45\textwidth]{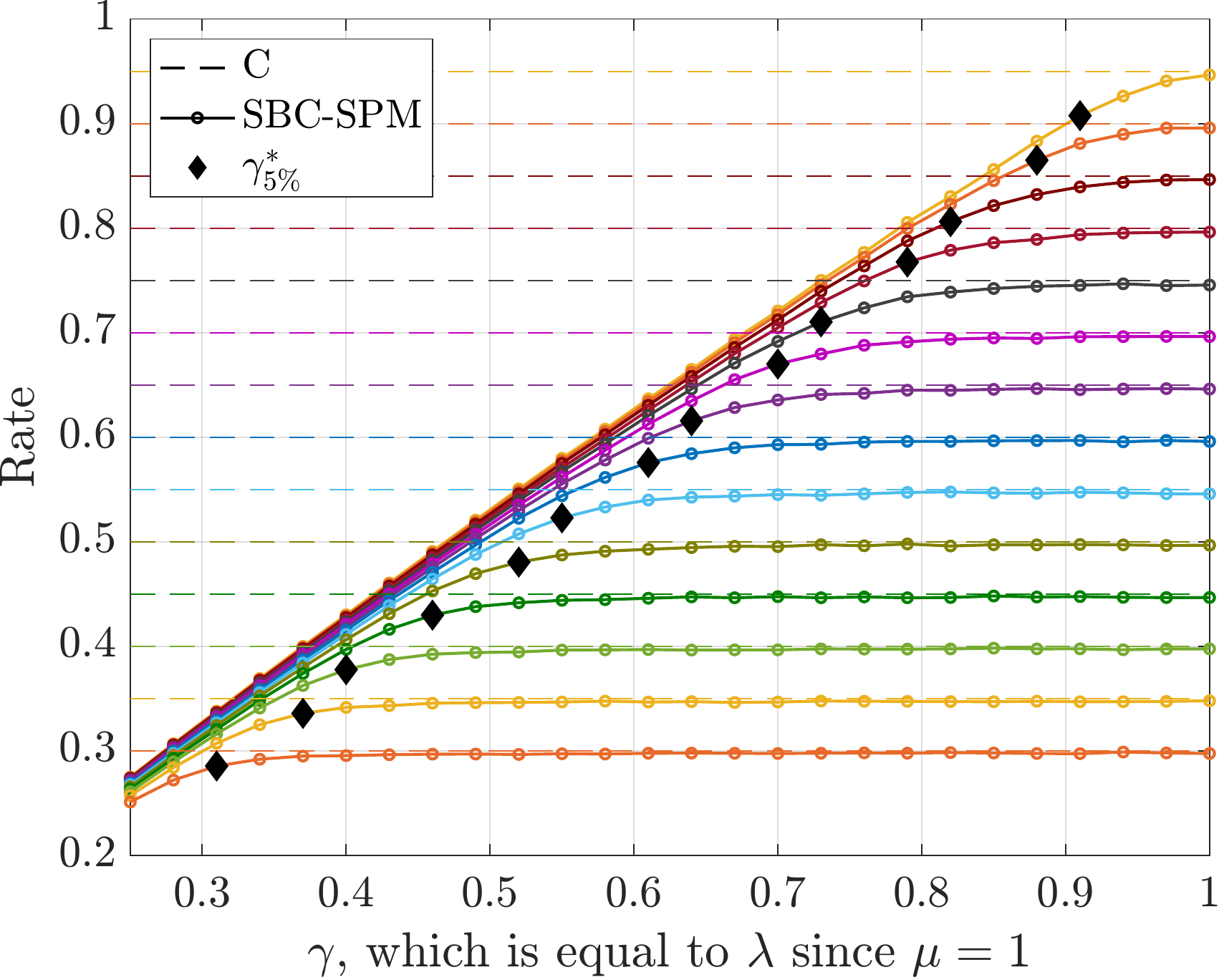}
\caption{Simulation results showing   rate $K/\tau$ vs $\gamma$ for SBC-SPM with $\mu = 1$.  The channel is BSC with capacities at intervals of 0.05, shown as dashed lines. The simulation results are averaged over 5000 trials with $K=240$ and  $N=8$ blocks with size $L_i = 30$. The black diamond markers show the operating points where the rate $K/\tau$ is within 5\% of the rate traditional SPM.}
\label{fig: rate vs gamma}
\end{figure}

Let $P(\theta \in S_0) = P_0$ and  $P(\theta \in S_1) = 1-P_0$. The distribution and expectation of $w_\theta$ for channel crossover probability $p$ with $q=1-p$ are as follows:
\begin{equation}
    \begin{aligned}
    w_\theta(t) &=
    \begin{cases}
        \frac{q}{P_0 q + (1-P_0) p} \quad w.p. \quad P_0 q \\
        \frac{p}{P_0 p + (1-P_0) q} \quad w.p. \quad  P_0 p \\
        \frac{p}{P_0 q + (1-P_0) p} \quad w.p. \quad (1-P_0) p \\
        \frac{q}{P_0 p + (1-P_0) q} \quad w.p. \quad (1-P_0) q \\
    \end{cases}
    \end{aligned}
\end{equation}
\begin{equation}
    \begin{aligned}
    E[w_\theta(t)] &=
        q \frac{P_0q}{P_0 q + (1-P_0) p} + 
        p \frac{P_0p}{P_0 p + (1-P_0) q} + \\
        p &\frac{(1-P_0)p}{P_0 q + (1-P_0) p}+
        q \frac{(1-P_0)q}{P_0 p + (1-P_0) q} \, .\\
    \end{aligned}
    \label{eq:Ew_theta_t}
\end{equation}

All denominators on the right side of \eqref{eq:Ew_theta_t} are positive when $p \in (0,\frac{1}{2})$ and $P_0 \in [0, 1]$. Therefore, $E[w_\theta(t)]$ is a continuous, symmetric and differentiable function of $p, P_0$. As shown by the equations below, the second derivative of $E[w_\theta(t)]$ is negative so that $E[w_\theta(t)]$ is concave in $P_0$.
The middle equation shows that $P_0 = \frac{1}{2}$ is the maximizer of $E[w_\theta(t)]$.
Thus, these equations prove that the single point-wise maximum of $E[w_\theta(t)]$ is achieved  when the sets $S_0$ and $S_1$ are equal in probability.
\begin{align*}
    &\frac{d}{d P_0}E[w_\theta(t)]
        =\frac{pq(q-p)}{(P_0 (q-p) + p)^2}
        -\frac{pq(q-p)}{(P_0 (q-p) - q)^2}
    \\
    &\frac{d}{d P_0} E[w_\theta(t)]\underset{P_0 = \frac{1}{2}}{\big|}
        =\frac{qp(q-p)}{(\frac{1}{2}(q+p))^2}
        -\frac{qp(q-p)}{(-\frac{1}{2}(q+p))^2} = 0
    \\
    &\frac{d^2}{d P_0^2}E[w_\theta(t)]
        =\frac{-2pq(q-p)^2}{(P_0 (q-p) + p)^3}
        +\frac{2pq(q-p)^2}{(P_0 (q-p) - q)^3}
\end{align*}

Once $P(\hat{B}^\theta_i=B^\theta_i) > \frac{1}{2}$, it becomes impossible to maximize $E[\frac{\rho_\theta(t+1)}{\rho_\theta(t)}]$ by partitioning the sub-block sequences into approximately equally likely $S_0$ and $S_1$.   By combining the sub-blocks, the transmitter is able to partition the overall sets $S_0$ and $S_1$ so that they are equally likely until the last few transmissions.

Furthermore, as $P(\hat{B}^\theta_i=B^\theta_i)$ increases further, continued transmissions have diminishing value. For example, with  $P(\hat{B}^\theta_i=B^\theta_i) >> \frac{1}{2}$, further transmissions that only consider that segment $B_i$ cannot contribute to increase the overall belief $\rho_\theta(\tau)$ beyond
$\Pi_{j\neq i} P(\hat{B}_i = B^\theta_i)$ are therefore inefficient.

Thus, overall efficiency demands an algorithm that combines sub-blocks before their individual posteriors are too large, preferably before $P(\hat{B}^\theta_i=B^\theta_i) > \frac{1}{2}$.  We will see that as $\gamma$ decreases, it becomes more difficult to avoid inefficient transmissions that fail to maximize $E[\frac{\rho_\theta(t+1)}{\rho_\theta(t)}]$.  However, these low-$\gamma$ cases often are also cases where the decoding time $T_d$ is already close to the minimum of $K/\lambda$.

\begin{figure}[H]
\centering
\includegraphics[width=0.45\textwidth]{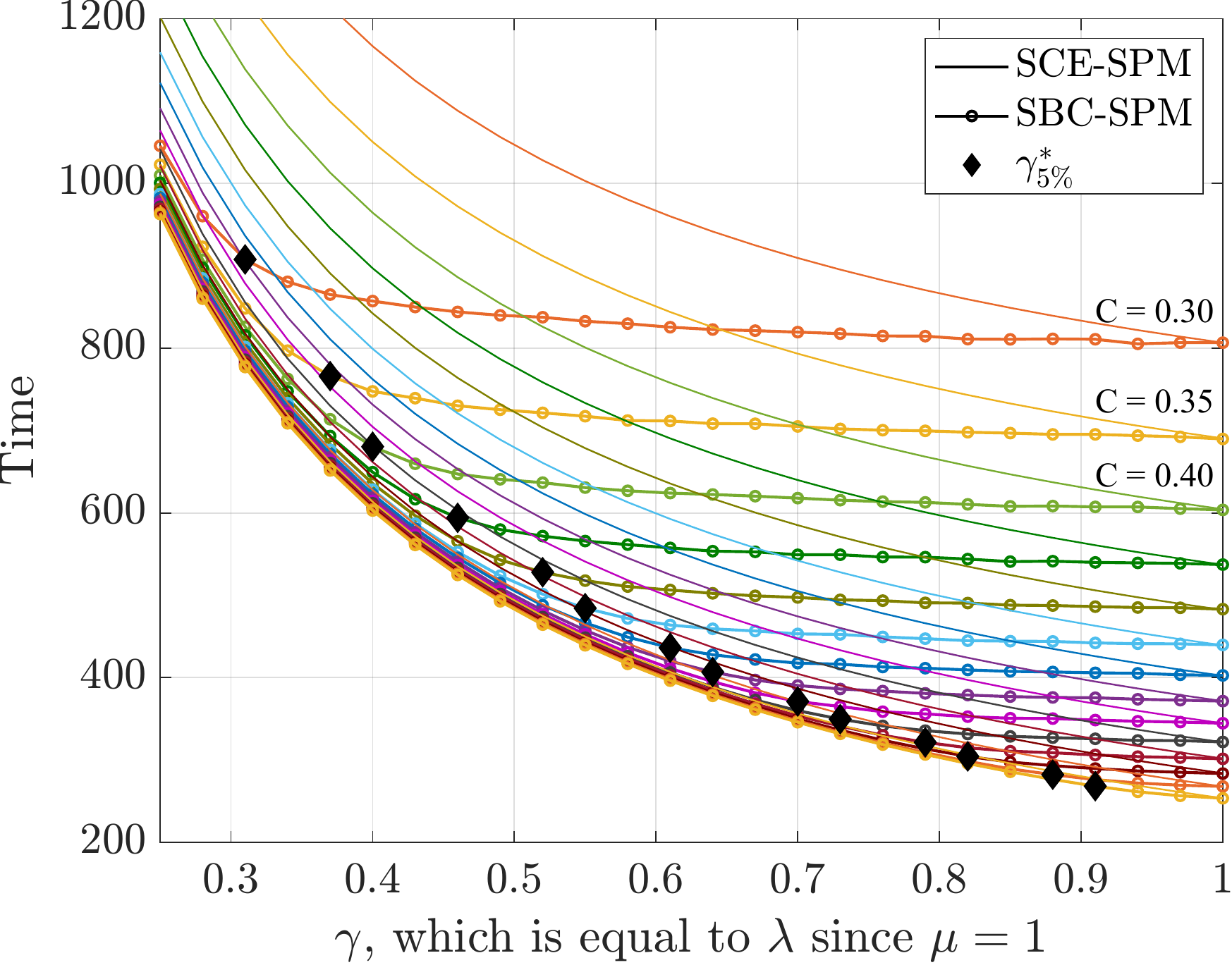}
\caption{Simulation results for time of first successful decoding $T_d$ for SBC-SPM (solid with dots) and SCE-SPM (thin solid) as a function of $\gamma$ for BSCs with various capacities indicated by the different colors as in Fig 4. The simulation results are averaged over 5000 trials with $K=240$ and  $N=8$ blocks with size $L_i = 30$. The black diamond markers show the operating points where the rate $K/\tau$ is within 5\% of the rate traditional SPM.}
\label{fig: time vs gamma}
\end{figure}
\section{Simulation Results}
\label{sec: simulation}
This section presents C++ simulation results for SBC-SPM on BSC channels with capacities ranging from 0.3 to 0.95 for $K=240$, $N=8$ and $0.5\le \gamma$.  

Fig. \ref{fig: rate vs gamma} shows the rate $K/\tau$ of the SBC-SPM algorithm as a function of $\gamma$ for  BSC channels with  capacities indicated by different colors.  The diamonds show the $\gamma$ values where the SBC-SPM algorithm attains a rate within 5\% of the SPM algorithm.  Observe that rates can approach capacity even for values of $\gamma < 1$, especially for lower capacity channels, where the expected number of transmissions is much larger than the message length. 

Fig. \ref{fig: time vs gamma} shows the time $T_d$ at which SBC-SPM decoding completes (solid lines with dots) as a function of $\gamma$ for the cases simulated in Fig. \ref{fig: rate vs gamma}.  Fig. 5 also shows the time $T_d$ at which SCE-SPM  decoding completes (solid lines without dots) as a function of $\gamma$ for these cases. As with Fig. \ref{fig: rate vs gamma}, the diamonds in Fig. \ref{fig: time vs gamma} show the decoding times where the SBC-SPM algorithm attains a rate within 5\% of the SPM algorithm. Fig. \ref{fig: time vs gamma} shows that the SBC-SPM algorithm achieves much lower decoding times than SCE-SPM for most $\gamma$ values, converging monotonically to SCE-SPM as $\gamma \rightarrow 1$. 


For each BSC, most of the decrease in decoding time offered by Causal Encoding, such as SBC-SPM, happens by the $\gamma$ value where the rate begins to transition away from capacity, i.e. at the black diamonds. For these and larger $\gamma$ values the transmitted symbols can be used effectively. For smaller $\gamma$ values, more symbols are available before the message fully arrives than can be used effectively, causing the rate to fall below capacity.  For larger values of $\gamma$, the message arrives so quickly that the benefit of causal encoding over traditional encoding is limited.

\begin{figure}[t]
\centering
\includegraphics[width=0.45\textwidth]{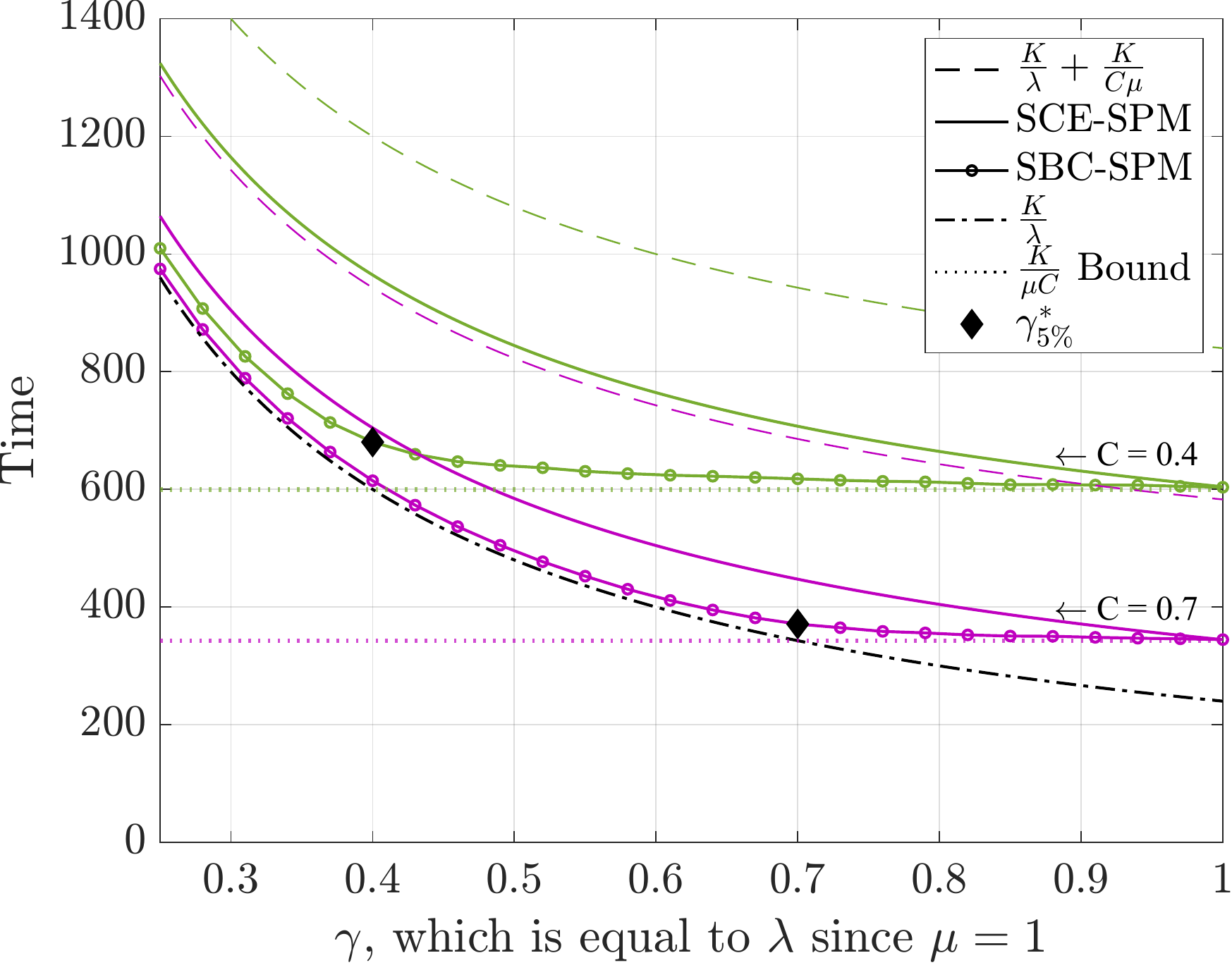}
\caption{This figure compares the SBC-SPM of Fig. 5 (solid with dots) to three bounds for the two example capacities of 0.4 and 0.7.  The time $K/\lambda$ at which the transmitter has the full message is show as a black dash-dot line. The lower bound $K/\lambda + K/C\mu$  on traditional communication is shown as a dashed line. Dot Constant line is the lower bound $K/\lambda + (K-KC)/C\mu$ on the expected decoding time $E[T_d]$ set by the link rate and channel capacity. Solid lines without dots are the data obtain ins simulations of the SCE-PM algorithm.}
\label{fig: time vs rate}
\end{figure}

Figure \ref{fig: time vs rate} again shows $T_d$ vs. $\gamma$ for SBC-SPM and SCE-SPM, but focuses on two BSC channels to compare with the analytical curves $K/\lambda$ and $K/\lambda + K/C\mu$.  The curve $K/\lambda$ indicates the time at which the message has fully arrived at the transmitter.  No algorithm can complete decoding before this time.  We see that for low $\gamma$ values, SBC-SPM performs close to this limit.  The curve $K/\lambda + K/C\mu$ shows the lower bound on decoding time for a traditional encoder that does not begin sending symbols until the message has fully arrived.  This bound assumes the transmitter can send reliably as a rate equal to capacity.  This curve shows that there is significant benefit in $T_d$ provided by causal encoding at any value of $\gamma$.

While these figures only show simulations with K=240 and N = 8, larger SBC-SPM simulations with much larger values of K, up to 16,000 have be successfully completed, 
which is beyond what SPM simulations in \cite{9174232} were able to achieve. Breaking the message into sub-blocks, even when not necessary for causal encoding, lowers complexity and makes the implementation more robust to numerical instability.



\section{Conclusion}
This paper introduces the SBC-SPM algorithm for low complexity causal encoding over the BSC.  Breaking the message into sub-blocks enables efficient causal encoding  without sacrificing the performance of PM. The algorithm elegantly groups collections of messages with equal and different posterior probabilities into binary trees and lists to manage complexity.  We show that equally-likely signaling maximizes efficiency and explore conditions when this balance can and cannot be approximated. We have also described the regions of $\gamma=\lambda/\mu$ where the new algorithm presents an advantage over non-causal and systematically causal algorithms, and the region where we can attain the capacity-approaching rate performance of PM. The new SBC-SPM algorithm provides implementation advantages even when causal encoding is not a requirement.
\label{sec: conclusion}

\bibliographystyle{IEEEtran}
\bibliography{IEEEabrv,references}

\end{document}